\documentclass[10pt,conference]{IEEEtran}
\IEEEoverridecommandlockouts
\usepackage{cite}
\usepackage{amsmath,amssymb,amsfonts}
\usepackage{algorithmic}
\usepackage{graphicx}
\usepackage{textcomp}
\usepackage{xcolor}
\usepackage{hyperref}
\usepackage{subcaption}
\usepackage{balance}

\def\BibTeX{{\rm B\kern-.05em{\sc i\kern-.025em b}\kern-.08em
    T\kern-.1667em\lower.7ex\hbox{E}\kern-.125emX}}
\begin{document}

\makeatletter

\newcommand{\newlineauthors}{%
\end{@IEEEauthorhalign}\hfill\mbox{}\par
\mbox{}\begin{@IEEEauthorhalign}}  

\makeatother


\title{Towards a Classification of Open-Source ML Models and Datasets for Software Engineering}

\author{\IEEEauthorblockN{Alexandra González}
\IEEEauthorblockA{\textit{Universitat Politècnica de Catalunya}\\
Barcelona, Spain \\
alexandra.gonzalez.alvarez@upc.edu}
\and
\IEEEauthorblockN{Xavier Franch}
\IEEEauthorblockA{\textit{Universitat Politècnica de Catalunya}\\
Barcelona, Spain \\
xavier.franch@upc.edu}
\newlineauthors 
\and
\IEEEauthorblockN{David Lo}
\IEEEauthorblockA{\textit{Singapore Management University}\\
Singapore \\
silverio.martinez@upc.edu}
\and
\IEEEauthorblockN{Silverio Martínez-Fernández}
\IEEEauthorblockA{\textit{Universitat Politècnica de Catalunya}\\
Barcelona, Spain \\
silverio.martinez@upc.edu}
}


\maketitle

\begin{abstract}  
Background: Open-Source Pre-Trained Models (PTMs) and datasets provide extensive resources for various Machine Learning (ML) tasks, yet these resources lack a classification tailored to Software Engineering (SE) needs. 
Aims: We apply an SE-oriented classification to PTMs and datasets on a popular open-source ML repository, Hugging Face (HF), and analyze the evolution of PTMs over time.
Method: We conducted a repository mining study. We started with a systematically gathered database of PTMs and datasets from the HF API. Our selection was refined by analyzing model and dataset cards and metadata, such as tags, and confirming SE relevance using Gemini 1.5 Pro. All analyses are replicable, with a publicly accessible replication package.
Results: The most common SE task among PTMs and datasets is \textit{code generation}, with a primary focus on \textit{software development} and limited attention to \textit{software management}. Popular PTMs and datasets mainly target \textit{software development}. Among ML tasks, \textit{text generation} is the most common in SE PTMs and datasets. There has been a marked increase in PTMs for SE since 2023 Q2.
Conclusions: This study underscores the need for broader task coverage to enhance the integration of ML within SE practices.
\end{abstract}


\section{Introduction}

The fast expansion of open-source platforms like Hugging Face (HF) \cite{huggingface} has enhanced access to Machine Learning (ML) models and datasets, driving advancements across various domains. With a consistent and significant uptrend in development activities on HF \cite{10.1145/3643991.3644898}, it is distinguished by its vast collection of Pre-Trained Models (PTMs), compared to other platforms \cite{10.1145/3569934}\cite{10.1145/3643991.3644907}. However, the categorization of these resources overlooks the specific needs of Software Engineering (SE). SE tasks frequently involve \textit{code generation}, \textit{code analysis}, and \textit{bug detection}, which differ significantly from the tasks commonly addressed by general-purpose ML models such as \textit{object detection} or \textit{image segmentation}. Therefore, the motivation for this work is to address this gap, as the absence of SE-specific categorization limits the efficient application of ML in SE tasks, potentially slowing down SE innovation. By providing a framework that aligns ML tasks with SE needs, this research aims to make the selection of PTMs and datasets more relevant and effective for SE practitioners and researchers, thus addressing a critical need within the field \cite{10.1145/3661167.3661215}.

The main contributions of this work are: (a) proposing and proving the feasibility of a preliminary classification framework for PTMs and datasets hosted on HF, tailored to SE needs; (b) providing advanced analysis, including the exploration of the relationship between SE activities and ML tasks, as well as the evolution of SE PTMs over time; (c) presenting a reproducible pipeline that accesses the HF API, filters, refines, and classifies resources on specific SE tasks.

\textbf{Data availability statement}: All research components, including the original and preprocessed data, along with all scripts for data collection, preparation, and analysis, are publicly available on Zenodo \cite{replication_package}. This ensures transparency and enables independent replication of the study, which is essential for updating the classification as new open-source PTMs and datasets are constantly being released.

\section{Related Work}

A systematic literature review conducted by Hou et al. \cite{10.1145/3695988} analyzed 395 research papers from January 2017 to January 2024 and categorized Large Language Models (LLMs) into SE tasks. These tasks were grouped into SE activities according to the six phases of the \textbf{Software Development Life Cycle}: \textit{requirements engineering}, \textit{software design}, \textit{software development}, \textit{software quality assurance}, \textit{software maintenance}, and \textit{software management}. 
Di Sipio et al. \cite{10.1145/3661167.3661215} highlighted the lack of a SE classification of PTMs on HF, as the existing one is specific for ML. To address this gap, they proposed extracting information from the model cards \cite{mitchell2019model} and using a semi-automated method to identify SE tasks and their corresponding PTMs from the literature. However, they only tested the mapping on three PTMs: \textit{BERT}, \textit{RoBERTa}, and \textit{T5}.
Yang et al. \cite{yang2024ecosystemlargelanguagemodels} analyzed the ecosystem of LLMs as of August 2023, curating 366 models and 73 datasets from HF for SE. The study also explores the use of LLMs to assist in constructing and analyzing the ecosystem, which increased the model size by 16.5. They focus on code-based LLMs, which represent a more specific scope compared to LLMs for broader SE tasks. 

Despite the above notable advances, a comprehensive framework that organizes PTMs and datasets on HF with a strong SE orientation remains absent. In contrast to prior works, this paper adopts a SE perspective to extend the existing taxonomy in Hou et al. \cite{10.1145/3695988}, to encompass a broader and more diverse set of PTMs, as well as datasets, categorizing them according to specific SE tasks and activities. By analyzing a substantially larger set of PTMs and SE-relevant datasets on HF, we address the unique SE-specific requirements unmet by previous studies, offering a novel, exhaustive preliminary classification framework for the community.

\section{Methodology}

\subsection{Research Objectives}
Following the Goal Question Metric (GQM) template \cite{caldiera1994goal}, our goal is to \textbf{\textit{analyze} PTMs and datasets \textit{for the purpose of} their classification \textit{with respect to} their application to SE tasks and activities \textit{from the point of view of} software engineers \textit{in the context of} the HF Hub}.

This goal is structured around four RQs. First, we need to assess the quality of model and dataset cards, as this information is essential for the subsequent RQs:

\begin{itemize}
    \item \textbf{RQ1}: What is the status of the model and dataset cards?
\end{itemize}

Next, we explore how the selected resources tackle SE:
\begin{itemize}
    \item \textbf{RQ2}: How do PTMs and datasets address SE?

    - \textbf{RQ2.1}: What SE tasks and activities are covered by PTMs and datasets?

    - \textbf{RQ2.2}: What are the most popular PTMs and datasets that address SE activities?
\end{itemize}

Following this, we explore the connection between ML tasks in HF's classification and SE activities:

\begin{itemize}
    \item \textbf{RQ3}: How are ML tasks related to SE activities?
\end{itemize}

Lastly, we examine the long-term relevance of our findings:
\begin{itemize}
    \item \textbf{RQ4}: How stable is this information over time?
\end{itemize}


\subsection{Data collection and Preparation}

\begin{figure}
    \centering
    \includegraphics[width=1\linewidth]{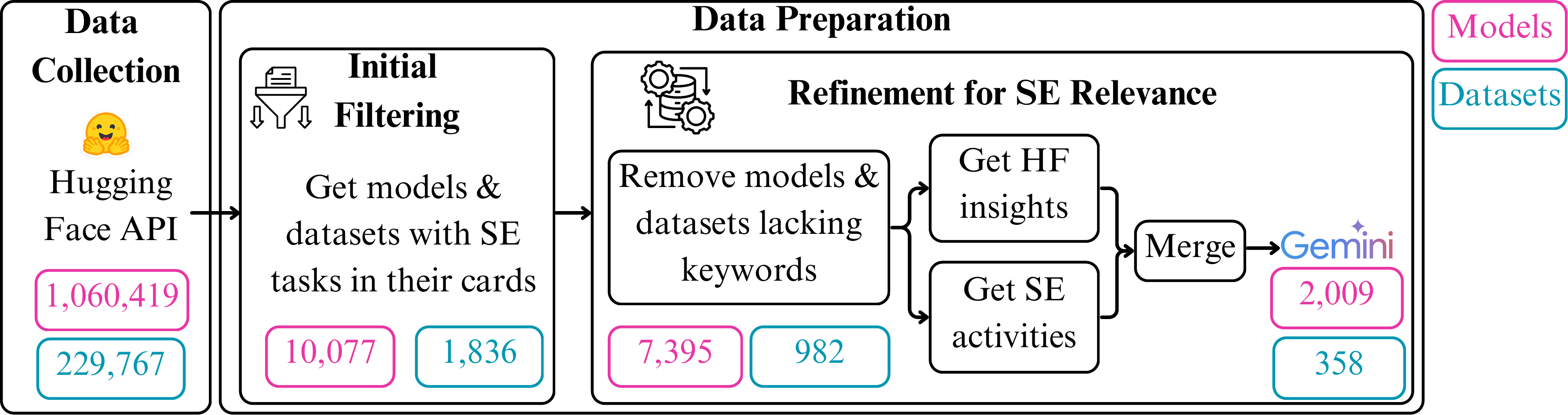}
    \caption{Data collection and preparation pipeline.} 
    \label{fig:pipeline}
\end{figure}

Figure \ref{fig:pipeline} illustrates the pipeline we followed to collect and prepare data for analysis. This process is designed for reproducibility, allowing anyone with access to the replication package \cite{replication_package} to validate and update the results. The inclusion criteria are summarized in Table \ref{tab:inclusion_criteria}.

\subsubsection{\textbf{Data Collection}} We used the HF API \cite{huggingfaceHuggingFace} to gather all available resources as of October 19, 2024: 1,060,419 PTMs and 229,767 datasets. During this process, we collected the unique identifiers for each model and dataset, and assessed the availability of their cards.

\subsubsection{\textbf{Data Preparation}} 
\paragraph{Initial Filtering} We automatically filtered PTMs and datasets mentioning SE tasks proposed by Hou et al. \cite{10.1145/3695988} in their cards. This step enabled us to focus on SE-relevant resources, resulting in 10,077 PTMs and 1,836 datasets.

\paragraph{Refinement for SE Relevance} To ensure our focus on SE, we removed entries that did not contain ``code'' or ``software'' in the model and dataset cards, resulting in 7,395 PTMS and 982 datasets. For this subset, we retrieved all available information from HF and mapped the SE task to the corresponding SE activity. Furthermore, we used an LLM, Gemini 1.5 Pro \cite{google_gemini}, to identify whether each resource was intended for SE based on an analysis of their cards and metadata. As a validation step, we manually classified a balanced sample of 30 PTMs and 30 datasets (between SE-relevant and non-SE), and compared our decisions with those generated by the LLM using Cohen's Kappa \cite{cohen1960coefficient}. This sample size was chosen to ensure reliable initial validation, as higher levels of agreement are anticipated \cite{sim2005kappa}. For PTMs, the Kappa was 0.80, which falls within the 0.61 to 0.80 range indicating substantial agreement \cite{landis1977measurement}; for datasets, it was 0.70, also reflecting this level of agreement. We noticed that some SE tasks, such as \textit{logging} and \textit{verification}, could be ambiguous. For instance, a model card containing a code snippet like \texttt{from transformers import logging} might be misclassified as a \textit{logging} task when it may not be. To ensure rigor, we asked the LLM to classify them by providing explicit definitions of such tasks. Lastly, we applied the prompts to all resources, resulting in 2,009 PTMs and 358 datasets, representing 0.19\% and 0.16\% of the original sets.

\begin{table}[!tb]
    \centering
    \caption{Inclusion criteria.}
    \label{tab:inclusion_criteria}
    \begin{tabular}{p{8.5cm}} 
    \hline
        \begin{enumerate}
            \item PTMs and datasets must be available on HF.
            \item PTMs and datasets must contain valid model and dataset cards.
            \item The cards must specify at least one SE task \cite{10.1145/3695988}.
            \item The cards must include ``code'' or ``software’’.
            \item An LLM (Gemini 1.5 Pro) must confirm relevance to SE.
        \end{enumerate} \\
    \hline
    \end{tabular} 
\end{table}

\subsubsection{\textbf{Data Analysis}}
In exploring the landscape of PTMs and datasets within HF, we centered our analysis on their cards, as well as on metadata such as tags, creation dates, and other relevant attributes. Additionally, we define popularity as the sum of the normalized number of likes and downloads.

\section{Results}
\subsection{What is the status of the model and dataset cards? (RQ1)}

As summarized in Table \ref{tab:mdls_readme}, over 33\% of the PTMs lack a model card, highlighting a gap in the documentation. Furthermore, more than 65\% do not mention any SE tasks, and only 0.95\% mention SE tasks. The analysis for datasets reveals a lack of documentation, as detailed in Table \ref{tab:ds_readme}. Over 27\% of datasets do not have a dataset card, indicating a significant gap in available information. Additionally, 0.56\% are empty, and a staggering 70.69\% do not reference any SE task. Only 0.80\% of the datasets allude to SE tasks.

\begin{table}
    \centering
    \caption{Summary of model cards.}
    \label{tab:mdls_readme}
    \begin{tabular}{ccc}
        \hline
        \textbf{Category} & \textbf{Number of PTMs}& \textbf{Proportion} \\
        \hline
        Not available & 350,524 & 33.05\% \\
        Available but empty & 3,686 & 0.35\% \\
        No SE tasks & 696,132 &  65.65\% \\
        With SE tasks & 10,077 &  0.95\%\\
        \hline
    \end{tabular}
\end{table}

\begin{table}
    \centering
    \caption{Summary of dataset cards.}
    \label{tab:ds_readme}
    \begin{tabular}{ccc}
        \hline
        \textbf{Category} & \textbf{Number of datasets}& \textbf{Proportion} \\
        \hline
        Not available & 64,210 & 27.95\% \\
        Available but empty & 1,294 & 0.56\% \\
        No SE tasks & 162,427 & 70.69\%\\
        With SE tasks & 1,836 & 0.80\%\\
        \hline
    \end{tabular}
\end{table}

\vspace{0.1cm}
\noindent
\fcolorbox{black}{white}{
    \parbox{0.95\linewidth}{
    \textbf{Finding 1}: The current state of HF shows that only 10,077 PTMs and 1,836 datasets mention SE tasks in their cards.
    }
}

\subsection{How do PTMs and datasets address SE? (RQ2)}

\subsubsection{\textbf{What SE tasks and activities are covered by PTMs and datasets?}}

Figure \ref{fig:mdls_barchart} shows the distribution of PTMs across SE tasks, color-coded by SE activity. The top tasks are \textit{code generation} and \textit{code completion}, with the first having nearly ten times as many PTMs as the third most common. The most represented activity is \textit{software development}, while \textit{software design} and \textit{requirements engineering} have limited PTMs. Additionally, the absence of PTMs for \textit{software management} highlights a critical gap in the current landscape.

\begin{figure}
    \centering
    \includegraphics[width=0.8\linewidth]{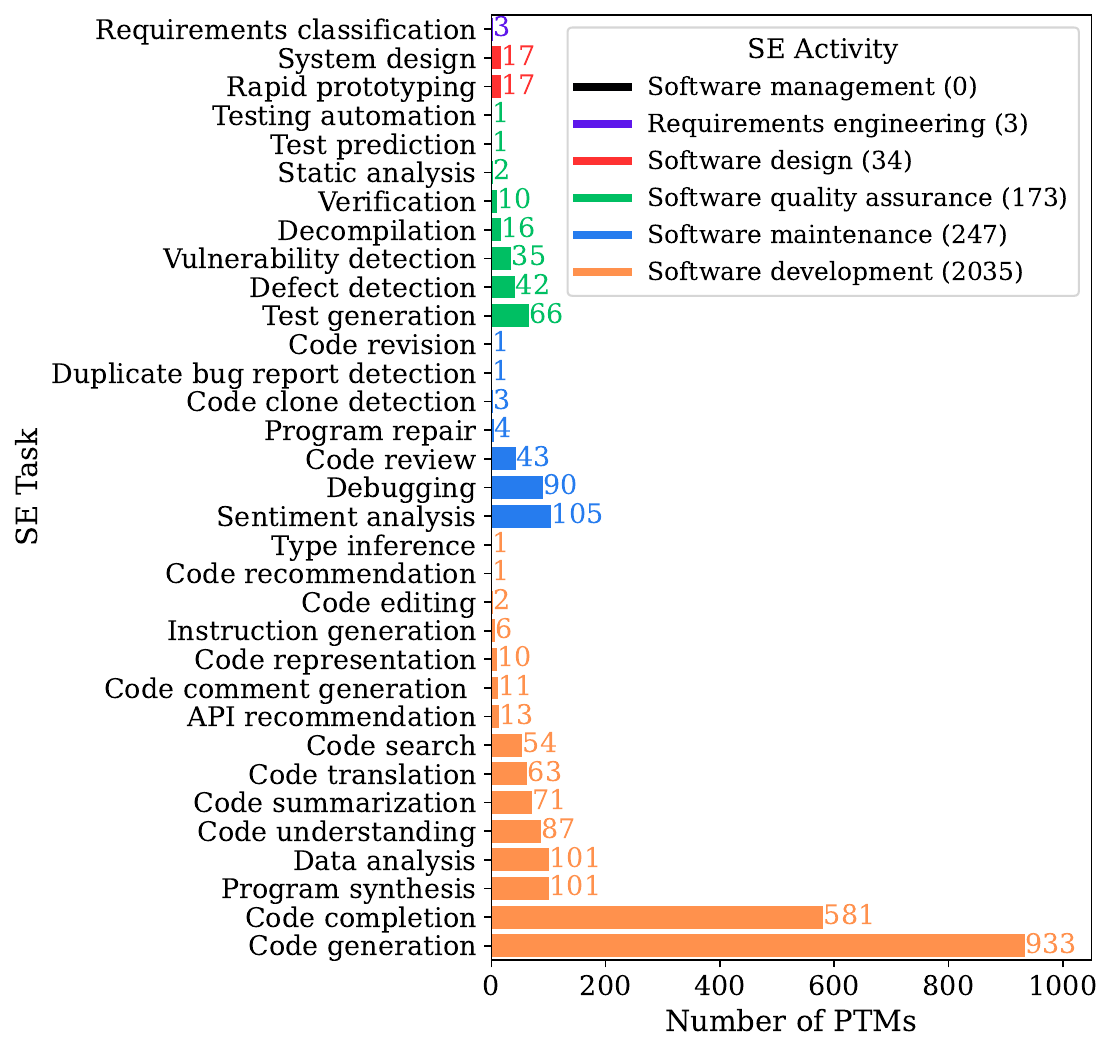}
    \caption{PTMs associated with each SE task and SE activity.}
    \label{fig:mdls_barchart}
\end{figure}

Regarding datasets, Figure \ref{fig:ds_barchart} shows that \textit{software development} dominates, while \textit{software design} and \textit{requirements engineering} are underrepresented, and \textit{software management} is entirely absent, mirroring the patterns observed with PTMs. The leading SE task is \textit{code generation}, which has over three times the number of datasets compared to the next task. 

Interestingly, the focus of secondary SE activity diverges between PTMs
and datasets: \textit{software maintenance} is the second SE activity
most addressed for PTMs, while for datasets, \textit{software quality assurance} holds this position. 

\begin{figure}
    \centering
    \includegraphics[width=0.8\linewidth]{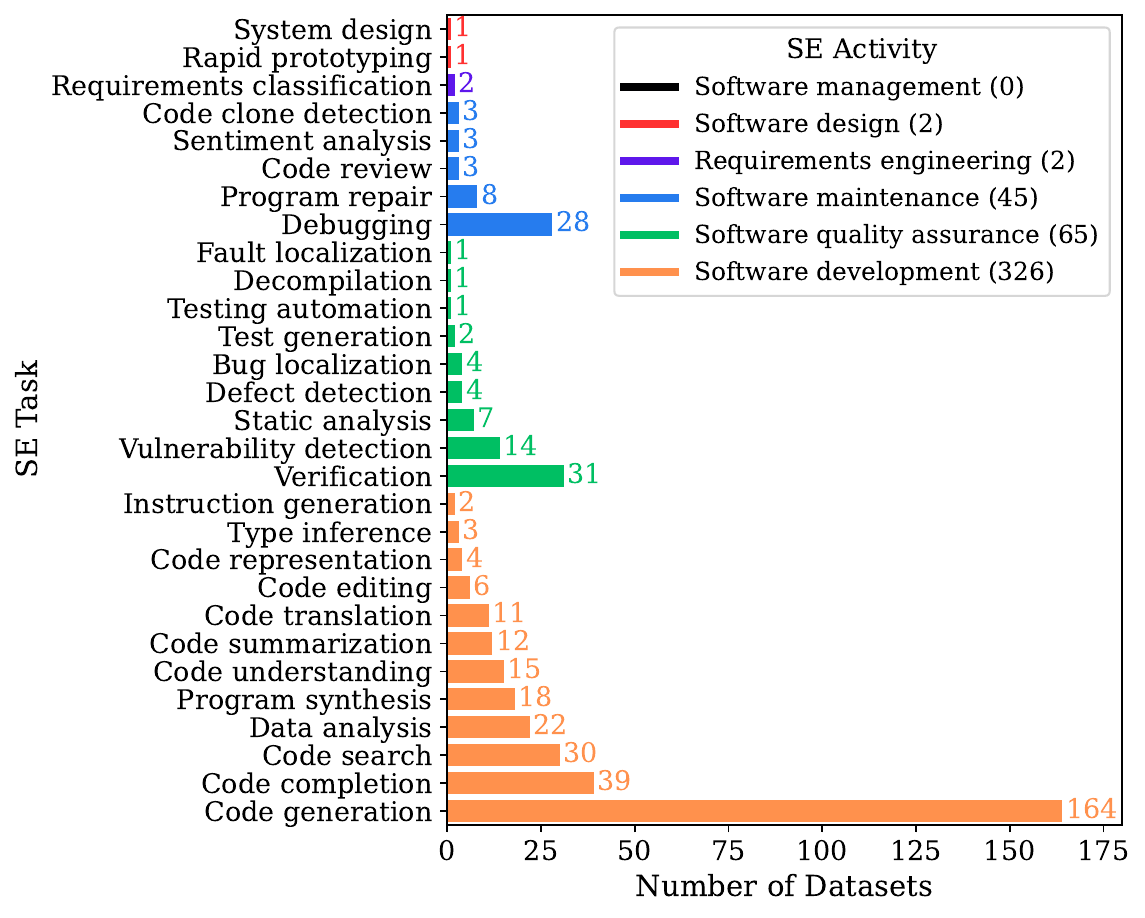}
    \caption{Datasets associated with each SE task and SE activity.}
    \label{fig:ds_barchart}
\end{figure}

\vspace{0.1cm}
\noindent
\fcolorbox{black}{white}{
    \parbox{0.95\linewidth}{
    \textbf{Finding 2.1}: \textit{Code generation} is the most covered SE task among PTMs and datasets.
    
    \textbf{Finding 2.2}: \textit{Software development} dominates, while \textit{software management} is absent in PTMs and datasets.
    }
}

\vspace{0.1cm}
\subsubsection{\textbf{What are the most popular PTMs and datasets that address SE activities?}}

As shown in Figure \ref{fig:mdls_top3_popularity}, the three most popular PTMs across SE activities reveal that \textit{software development} has a higher popularity due to its broad representation. Figure \ref{fig:ds_top3_popularity} presents the three most popular datasets for \textit{software development}, \textit{software quality assurance}, and \textit{software maintenance}. Other SE activities are omitted, as their dataset popularity is exactly 0. Popularity values for datasets tend to be higher, which may suggest that the community places more value on datasets than on PTMs, potentially indicating a higher demand for quality datasets in comparison to PTMs. In other words, researchers might perceive a good dataset as more valuable or essential for advancing SE activities than a well-performing model.

\begin{figure}
    \centering
    \includegraphics[width=0.8\linewidth]{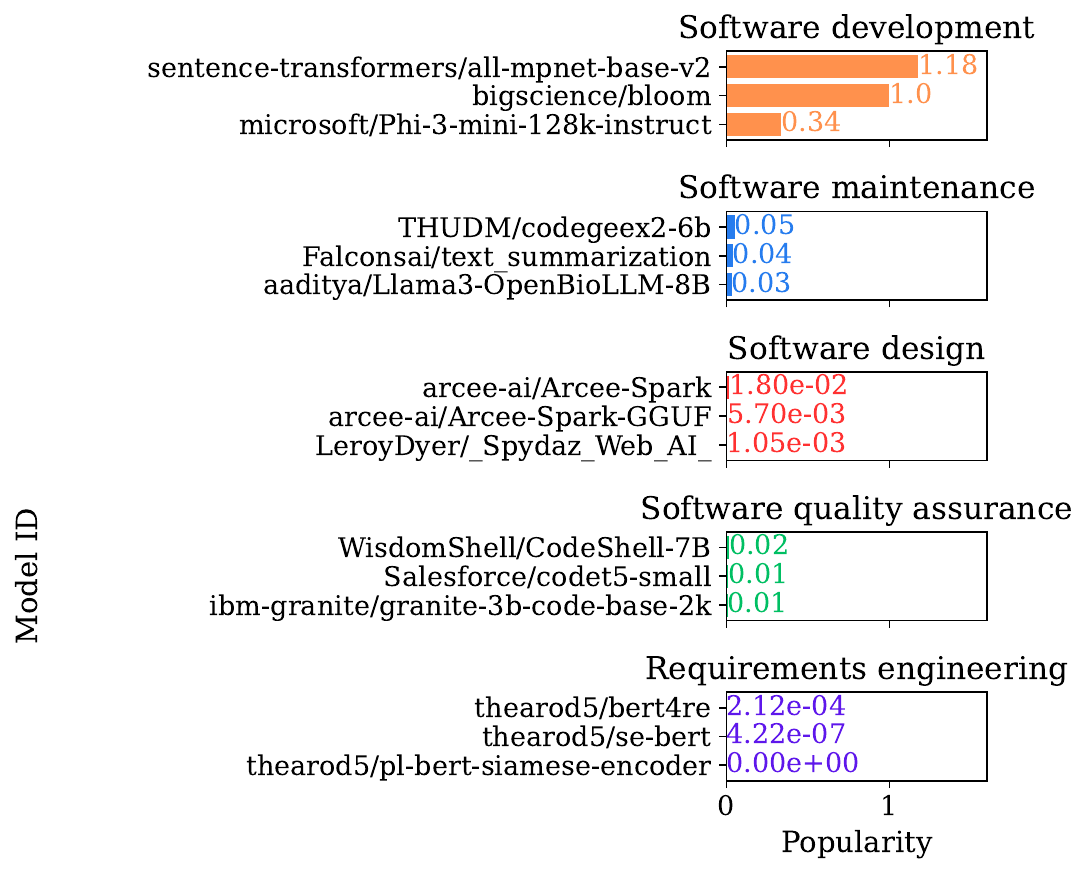}
    \caption{Top 3 most popular PTMs per SE activity.}
    \label{fig:mdls_top3_popularity}
\end{figure}

\begin{figure}
    \centering
    \includegraphics[width=0.8\linewidth]{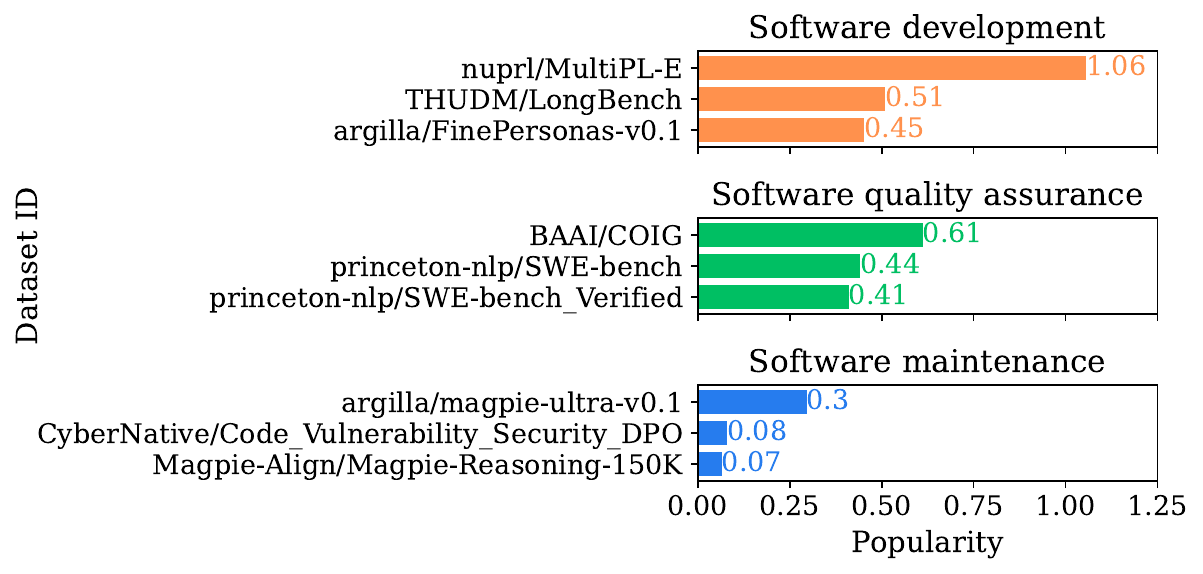}
    \caption{Top 3 most popular datasets per SE activity.}
    \label{fig:ds_top3_popularity}
\end{figure}

\vspace{0.1cm}
\noindent
\fcolorbox{black}{white}{
    \parbox{0.95\linewidth}{
    \textbf{Finding 2.3}: The most popular PTMs and datasets predominantly address \textit{software development}.
    
    \textbf{Finding 2.4}: Datasets tend to be more popular than PTMs, suggesting a higher demand for quality datasets in SE.
    }
}

\subsection{How are ML tasks related to SE activities? (RQ3)}

The relationship between SE activities and ML tasks for PTMs and datasets is illustrated in Figures \ref{fig:mdls_sankey} and \ref{fig:ds_sankey}, respectively, with the flow indicating the number of resources. Both figures highlight the prominence of the ML task \textit{text generation}, which is associated with the largest number of PTMs and datasets. This task is most commonly linked to \textit{software development} from the SE perspective, though it also supports other SE activities. In contrast, other ML tasks exhibit considerably smaller flows.

\begin{figure}
    \centering
    \includegraphics[width=0.85\linewidth]{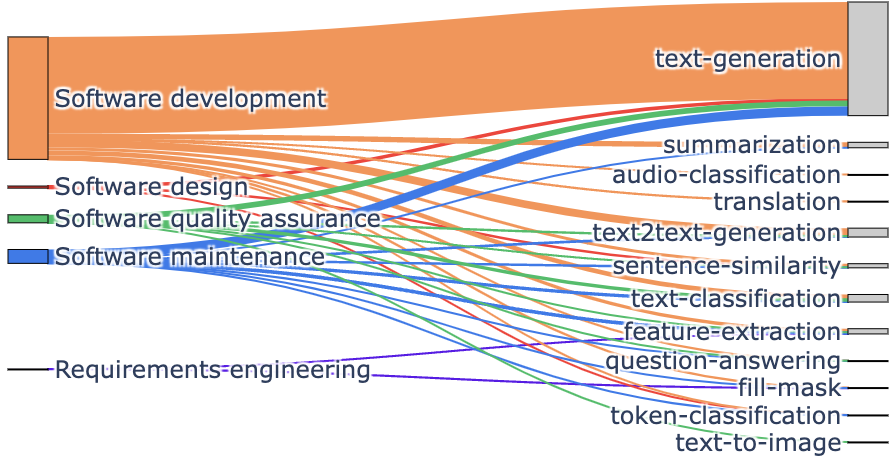}
    \caption{Association of SE activities with ML tasks for PTMs.}
    \label{fig:mdls_sankey}
\end{figure}

\begin{figure}
    \centering
    \includegraphics[width=0.85\linewidth]{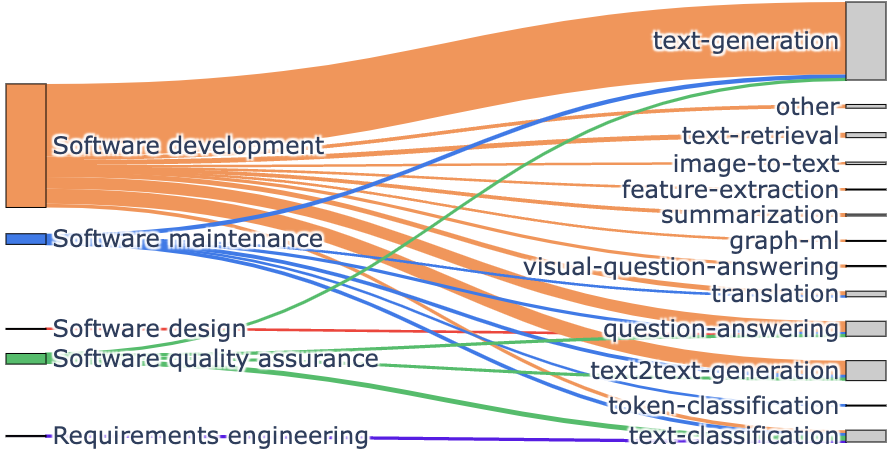}
    \caption{Association of SE activities with ML tasks for datasets.}
    \label{fig:ds_sankey}
\end{figure}

\vspace{0.1cm}
\noindent
\fcolorbox{black}{white}{
    \parbox{0.95\linewidth}{
        \textbf{Finding 3}: The ML task \textit{text generation} is the most common among SE-related PTMs and datasets.
    }
}

\subsection{How stable is this information over time? (RQ4)}

Figure \ref{fig:mdls_time_inset} shows a growth in the creation of PTMs for SE tasks since 2020, with quarterly data revealing periods of accelerated development. Notably, the percentage of PTMs for \textit{software development} increased from 6.78\% in 2023 Q2 to 20.55\% currently. Similarly, \textit{software maintenance} and \textit{software quality assurance} also showed significant growth, with the former rising from 1.70\% to 32.34\%, and the latter increasing from 1.19\% to 28.57\% over the same period. A zoom-in view of this period, from 2023 Q2 to 2024 Q3, is included in the figure to provide a closer examination. While the ranking of activities remains relatively stable over time, there are some fluctuations, particularly in 2023 Q3 and 2024 Q1, where \textit{software quality assurance} outperformed \textit{ software maintenance}, albeit with a less pronounced change in 2024. 

\begin{figure}
    \centering
    \includegraphics[width=1\linewidth]{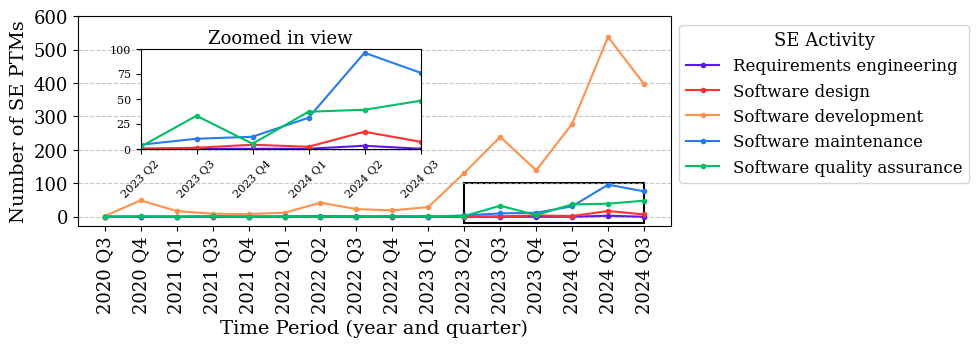}
    \caption{Number of SE PTMs created since 2020.}
    \label{fig:mdls_time_inset}
\end{figure}

\vspace{0.1cm}
\noindent
\fcolorbox{black}{white}{
    \parbox{0.95\linewidth}{
        \textbf{Finding 4.1}: PTMs for SE have grown consistently since 2020, with notable acceleration from 2023 Q2.
        
        \textbf{Finding 4.2}: The ranking of SE tasks remains stable over time, with minor fluctuations.
    }
}






\section{Discussion and Limitations}
Our results align with Hou et al. \cite{10.1145/3695988}, which shows a similar distribution of LLMs across SE activities, with both studies identifying an emphasis on \textit{code generation} tasks. However, we have found a gap in \textit{software maintenance} within the current state of HF. Notably, our findings complement Yang et al.'s analysis \cite{yang2024ecosystemlargelanguagemodels}, highlighting a rapid growth in this area. 

Potential threats impacting our study's validity are outlined. First, the quality of the model and dataset cards may compromise internal validity, as incomplete documentation could lead to misclassification. Our conclusions rely on the assumption that each PTM and dataset’s card accurately reflects the SE tasks it addresses. Second, other platforms/repositories (e.g., PyTorch Hub \cite{pytorchPyTorch}) may host additional relevant resources, and changes on HF could affect the broader applicability of our conclusions. Third, our classification of SE tasks and activities is based on a taxonomy from the existing literature \cite{10.1145/3695988}. Although this taxonomy informs our analysis, emerging standards may further enrich our understanding of SE. To mitigate these threats, our study is fully replicable, enabling future researchers to validate and extend our findings.

\section{Conclusion and Future Work}
This study examines the availability of PTMs and datasets for SE hosted on HF, revealing that almost 1\% reference SE tasks. PTMs predominantly focus on \textit{code generation} and \textit{code completion}, while \textit{code generation} is the most represented dataset task. SE activities like \textit{software design} and \textit{requirements engineering} are under-represented, with a gap in \textit{software management} resources. SE datasets tend to be more popular than PTMs. The most prevalent ML task across both PTMs and datasets is \textit{text generation}. Additionally, there has been a significant surge in PTMs for SE since 2023 Q2. This snapshot of the current landscape highlights the existing gaps in SE resources and provides valuable insight into the field's evolving needs, helping to motivate future research efforts aimed at addressing these underrepresented areas.

For future work, we plan to extend this classification to other repositories, such as Papers with Code \cite{paperswithcodePapersWith}, PyTorch Hub \cite{pytorchPyTorch} and TensorFlow Hub \cite{tensorflowTensorFlow}, allowing for a broader analysis of PTMs and datasets in SE.  In addition, we aim to refine the classification process by addressing current limitations, such as handling synonymous in SE tasks, to improve accuracy. To make our classification more actionable, we propose developing a dashboard to assist researchers with proper sampling and support practitioners in the SE community when selecting PTMs for real-world applications, thereby enhancing the relevance and impact of our work.

\section*{Acknowledgment}
This work has been funded by the Spanish research project DOGO4ML
(ref. PID202 0-117191RB-I00).

\balance
\bibliographystyle{IEEEtran}
\bibliography{references}

\end{document}